\documentstyle[12pt]{article}

\advance \topmargin by -\headheight
\advance \topmargin by -\headsep

\evensidemargin \oddsidemargin
\begin{document}
\font\ninerm = cmr9

\def\footnoterule{\kern-3pt \hrule width \hsize \kern2.5pt}

\pagestyle{empty}
\begin{center}
{\large\bf Black Hole Radiation (with and) without Weyl Anomaly}%
\footnote{\ninerm This work is supported
in part by funds provided by the U.S. Department of Energy (D.O.E.)
under cooperative agreement \#DE-FC02-94ER40818,
and by Istituto Nazionale di
Fisica Nucleare (INFN, Frascati, Italy). }

\vskip 1cm
G. Amelino-Camelia\footnote{Present
address: Theoretical Physics,
Oxford University, 1 Keble Rd., Oxford OX1 3NP, U.K.} and D. Seminara
\vskip 0.5cm
{\it Center for Theoretical Physics\\
Laboratory for  Nuclear Science and Department of Physics\\
Building 6, Massachusetts Institute of Technology\\
Cambridge, Massachusetts 02139, U.S.A.}

\end{center}

\vspace{1.2cm}
\begin{center}
{\bf ABSTRACT}
\end{center}

{\leftskip=0.6in \rightskip=0.6in

In the semiclassical analysis of black hole radiation in matter-coupled
dilaton gravity, a one-parameter ``$k$"-family of measures
for the path integral
quantization of the matter fields is considered.
The Weyl anomaly is proportional to the parameter $k$, but the
black hole radiation seen by minkowskian observers at
future null
infinity
is $k$-independent.
}

\vskip 3cm
\centerline{Submitted to: Classical and Quantum Gravity}

\vfill

\hbox to \hsize{MIT-CTP-2443 \hfil }

\newpage
\baselineskip 24pt plus .5pt minus .5pt
\pagenumbering{arabic}
\pagestyle{plain}

\section{Introduction}
In the conventional semiclassical analysis\cite{bhlect} of
black hole radiation\cite{haw75}
in matter-coupled dilaton gravity (MCDG)
\begin{eqnarray}
S_D = \int d^2 x \sqrt{-g} e^{-2 \phi} \left[
  R + 4 (\nabla \phi)^2 + 4 \lambda^2 \right]
- {1 \over 2} \sum_{n=1}^N \int d^2 x \sqrt{-g}
g^{\mu \nu} \partial_\mu f_n \partial_\nu f_n
{}~,
\label{sdilaton}
\end{eqnarray}
where $g$, $\phi$, and $f_n$ are
the metric, dilaton, and matter fields respectively,
a central role\cite{bhlect,chr77} is played by the Weyl anomaly.
One starts by treating
classically the gravitational collapse, and then
the matter degrees of freedom are quantized in
the background of the resulting black hole metric.
The Weyl anomaly
\begin{eqnarray}
g^{\mu \nu} T_{\mu \nu} = {N \over 24 \pi} R(g) ~,
\label{anopw}
\end{eqnarray}
where $T_{\mu \nu}$ is the
matter energy-momentum tensor\cite{bos}
\begin{eqnarray}
T_{\mu \nu} \equiv \sum_{n=1}^N < \partial_\mu f_n \partial_\nu f_n
- {1 \over 2} g_{\mu \nu} g^{\alpha \beta}
\partial_\alpha f_n \partial_\beta f_n >_g ~,
\label{tjack}
\end{eqnarray}
is a consequence of the diffeomorphism-invariant quantization of the matter
fields; for example, in the path integral formulation, Eq.(\ref{anopw})
follows from choosing the diffeomorphism-invariant
measure\cite{pol81}
\begin{eqnarray}
\int {\cal D}\delta f_n ~ {\rm exp}\left(i \int d^2 x ~ \sqrt{-g} ~
\delta  f_n \,  \delta f_n \right) ~ = ~ 1
{}~,\label{measp}
\end{eqnarray}
which is not Weyl-invariant.

Once appropriate physical boudary conditions are imposed,
Eq.(\ref{anopw}) and the covariant conservation of the
matter energy-momentum tensor
\begin{eqnarray}
\nabla_\mu (g^{\mu \nu} T_{\nu \alpha}) = 0  ~,
\label{anopd}
\end{eqnarray}
determine $T_{\mu \nu}$, which has three independent components in
1+1 dimensions, and therefore determine the black hole radiation.

The Weyl anomaly is usually considered\cite{bhlect,chr77}
to be a crucial ingredient of black hole radiation because
for a traceless matter energy-momentum tensor
Eq.(\ref{anopd}) would lead to no radiation.
In order to achieve a deeper understanding of
the relation between Weyl anomaly and black hole radiation,
in this Letter we study how
the conventional analysis of black hole radiation
in MCDG is affected by the modifications
to Eqs.(\ref{anopw}) and (\ref{anopd}) that arise
in the alternative
approaches to the quantization of the matter fields
which have been recently considered in
Refs.[5-9].

\section{$k$-dependent Anomaly Relations}
In the path integral formulation,
the alternative approaches to the quantization
of the matter fields
considered in Refs.\cite{bon86,neu90}
correspond to the choice of measure
\begin{eqnarray}
\int {\cal D}\delta f_n ~ {\rm exp}\left(i \int
d^2 x ~ (-g)^{k \over 2} ~
\delta  f_n \,  \delta f_n \right) ~ = ~ 1
{}~,\label{measb}
\end{eqnarray}
where $k$ is a fixed real parameter.

\noindent
We observe that this choice of measure
is invariant under infinitesimal variations of the
form\footnote{Note that the matter action in (\ref{sdilaton})
is invariant under diffeomorphisms and Weyl transformations
[and therefore in particular is invariant under the
transformations (\ref{ktrasf})]; this is related to the fact
that, if $g_{\mu \nu}$ is a density of weight $\rho$,
the transformations of $\sqrt{- g} g^{\mu \nu}$
under diffeomorphisms are independent of $\rho$.}
\begin{eqnarray}
\delta x^{\mu} \!=\! v^{\mu}
{}~,~~~~
\delta g_{\mu \nu}
\!=\! v^\alpha \partial_\alpha g_{\mu \nu}
\!+\! g_{\alpha \nu} \partial_\mu v^\alpha
\!+\! g_{\alpha \mu} \partial_\nu v^\alpha
\!+\! {1-k \over k} g_{\mu \nu} \partial_\alpha v^\alpha
{}~. \label{ktrasf}
\end{eqnarray}
In order to render this formula valid for all $k$'s,
we prescribe that
the singular limit $k \! \rightarrow \! 0$, which corresponds
to the theory considered
in Refs.[7-9],
be formally taken
so that $\partial_\mu v^\mu \! \rightarrow \! k w$
for $k \! \rightarrow \! 0$, where $w$
is an arbitrary function.
Following this limiting procedure,
at $k \! = \! 0$ the variations (\ref{ktrasf}) take the
form
\begin{eqnarray}
\delta x^{\mu} \!=\! v^{\mu}
{}~,~~~~
\delta g_{\mu \nu} \!=\! v^\alpha \partial_\alpha g_{\mu \nu}
\!+\! g_{\alpha \nu} \partial_\mu v^\alpha
\!+\! g_{\alpha \mu} \partial_\nu v^\alpha
\!+\! w g_{\mu \nu} ~,~~~~ {\rm with}~\partial_\mu v^{\mu} \! = \! 0
\label{wtrasf}
\end{eqnarray}
which indeed reproduce the invariance[7-9]
of the measure (\ref{measb}),
at $k \! = \! 0$,
under diffeomorphisms
of unit Jacobian ($\partial_\mu v^{\mu} \! = \! 0$) and
Weyl transformations.

Note that (\ref{ktrasf}) and (\ref{wtrasf}) indicate
that diffeomorphisms
of unit Jacobian
are a symmetry
of the measure (\ref{measb})
for every value of $k$.

As a result of the properties of the measure (\ref{measb}),
the matter energy-momentum tensor satisfies the following
anomaly relations
\begin{eqnarray}
g^{\mu \nu} T_{\mu \nu} \!&=&\! k {N \over 24 \pi} g^{{k-1 \over 2}}
R({\hat g})
{}~, \label{anobw}\\
\nabla_\mu (g^{\mu \nu} T_{\nu \alpha}) \!&=&\! (k-1) {N \over 48 \pi}
{1 \over \sqrt{-g}} \partial_\alpha [g^{{k \over 2}} R({\hat g})]
{}~,
\label{anobd}
\end{eqnarray}
where ${\hat g}_{\mu \nu} \equiv (-g)^{k-1 \over 2} g_{\mu \nu}$.
Obviously for $k \! = \! 1$ Eqs.(\ref{anobw}) and (\ref{anobd})
reproduce Eqs.(\ref{anopw}) and (\ref{anopd}), and
for $k \! = \! 0$ they reproduce the
corresponding relations encountered in Refs.\cite{kar94,jac95,abs95}.
For our investigation of the relation between Weyl anomaly
and black hole radiation, it is especially important
that (\ref{anobw})
is directly proportional to $k$, which, in particular, implies
that there is no Weyl anomaly in the $k \! = \! 0$ limit.

The invariance under the
transformations (\ref{ktrasf}) is encoded in the fact that
\begin{eqnarray}
\sqrt{-g} \, \nabla_\mu (g^{\mu \nu} T_{\nu \alpha}) +
{1-k \over 2 k} \partial_\alpha[\sqrt{-g} g^{\mu \nu} T_{\mu \nu}] = 0
{}~,
\label{relbsymm}
\end{eqnarray}
which is consistent with (\ref{anobw}) and (\ref{anobd}).

Interestingly, the relations (\ref{anobw}), (\ref{anobd}), and
(\ref{relbsymm}) can all be rewritten rather elegantly
in terms of ${\hat g}$
and $\hat\nabla$, the
covariant derivative computed with
the metric ${\hat g}$,
\begin{eqnarray}
{\hat g}^{\mu \nu} T_{\mu \nu} \!&=&\!  k {N \over 24 \pi}
R({\hat g}) ~,\label{anobhatw}\\
\hat\nabla_\mu ({\hat g}^{\mu \nu} T_{\nu \alpha}) \!&=&\!
(k-1) {N \over 48 \pi} \partial_\alpha R({\hat g})
{}~,
\label{anobhatd}\\
\hat\nabla_\mu ({\hat g}^{\mu \nu} T_{\nu \alpha}) \!\!&+&\!\!
{1-k \over 2k} \partial_\alpha ({\hat g}^{\mu \nu} T_{\mu \nu}) = 0
{}~.
\label{hatrelbsymm}
\end{eqnarray}
This is a consequence\cite{ags} of the fact that
the transformations (\ref{ktrasf})
can be obtained as the realization of the diffeomorphism
group on a tensorial density of weight ${1-k \over k}$.
(N.B.: $g_{\mu \nu}$ has weight ${1-k \over k}$ with respect
to the metric ${\hat g}_{\mu \nu}$.)

The fact  that the right-hand
sides of Eqs.(\ref{anobw})
and (\ref{anobd}) [or equivalently (\ref{anobhatw})
and (\ref{anobhatd})]
do not transform covariantly under general diffeomorphisms
implies that $T_{\mu \nu}$ is not a tensor.
One can show that,
under a coordinate
redefinition $x^\mu \rightarrow y^\mu$,
$T_{\mu \nu}$ transforms as follows
\begin{eqnarray}
T^{(x)}_{\mu \nu} \rightarrow T^{(y)}_{\mu \nu} =
(T^{(x)}_{\alpha \beta} + \Delta^{(x,y)}_{\alpha \beta})
{dx^\alpha \over dy^\mu}
{dx^\beta \over dy^\nu}
\label{ttrasf}
\end{eqnarray}
where
\begin{eqnarray}
\Delta^{(x,y)}_{\alpha \beta} \! = \!
\frac{N (\!1\!-\! k\!)}{24\pi}
\left ( \nabla_\alpha \! \nabla_\beta \! \ln \! J
\! - \! g_{\alpha \beta} \Box \! \ln \! J \right )+
\frac{N (\!1\!-\! k\!)^2}{96\pi}
( g_{\alpha \beta} \! \nabla_\gamma \! \ln \! J
\nabla^\gamma \! \ln \! J
\! - \! 2 \nabla_\alpha \! \ln \! J
\nabla_\beta \! \ln \! J \nonumber\\
+ \nabla_\alpha \! \ln \! \sqrt{-g} \,
\nabla_\beta \! \ln \! J \!
+ \! \nabla_\beta \! \ln \! \sqrt{-g} \, \nabla_\alpha \! \ln \! J \!
- \! g_{\alpha \beta} \nabla^\gamma \! \ln \! \sqrt{-g} \,
\nabla_\gamma \! \ln \! J
\! + \! g_{\alpha \beta} \Box \! \ln \! J )
\, ,
\label{deltamunu}
\end{eqnarray}
and $J$ is the Jacobian of
the transformation $x^\mu \rightarrow y^\mu$.

\noindent
Notice
that $\Delta^{(x,y)}_{\alpha \beta} \! = \!0$
whenever $J$ is constant.
This implies that
$T_{\mu \nu}$ transforms covariantly not only under
diffeomorphisms of unit Jacobian,
but also under dilatations ($y^\mu \! = \! c x^\mu$
with constant $c$); in fact,
a general
coordinate redefinition of constant Jacobian
can be obtained as the
composition of a diffeomorphism of unit Jacobian
and a dilatation.
The covariance of $T_{\mu \nu}$ under this
diffeomorphism subgroup which is
larger than the one leaving invariant the measure (\ref{measb})
can be understood\cite{ags}
as a consequence of the fact that in 1+1 dimensions
$\sqrt{-g} R(g)$ is a total derivative\cite{jac95}.

\section{Black Hole Radiation}
\subsection{Conformal-Gauge Analysis for $k \! = \! 1$}
We now turn to the study of black hole radiation,
starting with a brief review of the conventional ($k \! = \! 1$)
approach to the problem,
{\it i.e.} assuming that the
relations (\ref{anopw}) and (\ref{anopd}) hold.
For simplicity, we limit our analysis to the example of black hole
discussed in Ref.\cite{bhlect}, and keep
our notation consistent with the one of Ref.\cite{bhlect};
in particular, we
introduce light-cone coordinates $\sigma^\pm$
and work in conformal gauge: $g_{+-} \! = \! - e^{2 \rho}/2$,
$g_{++} \! = \! g_{--} \! = \! 0$.

The black hole
is formed by collapse of a shock-wave,
traveling in the $\sigma^-$ direction, described by the stress
tensor\footnote{We choose to write the magnitude of the shock-wave as
$a e^{-\sigma^+_0}$, in order to keep our notation ``$a$"
consistent with the one of Ref.\cite{bhlect}.}
\begin{eqnarray}
{1 \over 2}\partial_+ f\partial_+ f=
a e^{-\sigma^+_0} \delta (\sigma^+ - \sigma^+_0)
{}~.
\label{shock}
\end{eqnarray}
The solution of the MCDG classical equations of motion,
taking into account that
for $\sigma^+ \! < \! \sigma^+_0$
we are in the vacuum, gives a
black hole background metric with conformal factor
\begin{eqnarray}
\rho = - {1 \over 2}
\ln \left[1+ \Theta(\sigma^+ - \sigma^+_0)
{a \over \lambda} e^{\lambda \sigma^-}
\left( e^{\lambda (\sigma^+_0-\sigma^+)}-1 \right) \right]
{}~.
\label{background}
\end{eqnarray}

The next step in the
conventional semiclassical
analysis of this black hole, consists in
using the quantum relations (\ref{anopw}) and (\ref{anopd}) to derive
the flux of matter energy across\footnote{Like in Ref.\cite{bhlect},
${\cal I}_R^+$ (${\cal I}_R^-$) is
the future (past) null
infinity for right-moving light rays, and analogously
${\cal I}_L^+$ (${\cal I}_L^-$) is
the future (past) null
infinity for left-moving light rays.}
${\cal I}_R^+$, which is given by the value
of $T_{--}$ on ${\cal I}_R^+$.
In conformal gauge (\ref{anopw}) and (\ref{anopd}) take the form
\begin{eqnarray}
& T_{+-} = - {N \over 12 \pi} \partial_+ \partial_- \rho
{}~,&
\label{stroconfw} \\
& \partial_\mp T_{\pm \pm}+ \partial_\pm T_{+-}
- 2  T_{+ -} \, \partial_\pm \rho =0
{}~,&
\label{stroconfd}
\end{eqnarray}
and these lead to
\begin{eqnarray}
T_{\pm \pm} = {N \over 12 \pi}
\left[ \partial^2_\pm \rho -
(\partial_\pm \rho)^2 + t_\pm(\sigma^\pm) \right] ~.
\label{solstroconfd}
\end{eqnarray}
The functions of integration $t_\pm$  are to be determined
by imposing physical boundary conditions, which for our
collapsing shock-wave consist\cite{bhlect} in requiring
that $T_{\mu \nu}$  vanish
on ${\cal I}_L^-$ ({\it i.e.} $\sigma^+ \! = \! - \infty$),
and that there be no incoming radiation
along ${\cal I}_R^-$ ({\it i.e.} $\sigma^- \! = \! - \infty$) except
for the classical shock-wave at $\sigma^+ \! = \! \sigma^+_0$;
this implies that $t_\pm \! = \! 0$.
Substituting $t_\pm \! = \! 0$ and (\ref{background}) in
(\ref{stroconfw}) and (\ref{solstroconfd}) one easily derives that
on ${\cal I}^+_R$ ({\it i.e.} $\sigma^+ \rightarrow \infty$)
\begin{eqnarray}
\left[T_{++}\right]_{{\cal I}^+_R} = 0 ~,~~~~
\left[T_{+-}\right]_{{\cal I}^+_R} = 0 ~,~~~~~~~~
\label{confstrozerosigma}\\
\left[T_{--}\right]_{{\cal I}^+_R}
= { N a \lambda^2 \over 48 \pi} e^{\lambda \sigma^-}
{2 \lambda - a e^{\lambda \sigma^-}
\over \left(\lambda - a e^{\lambda \sigma^-} \right)^2}
{}~.
\label{confstrommsigma}
\end{eqnarray}

As clarified in Ref.\cite{bhlect},
the physical interpretation of this solutions is
clearest
in the $y^\pm$ coordinates
\begin{eqnarray}
y^+ = \sigma^+~,~~~ y^- = - \ln \left(e^{- \lambda \sigma^-} -
a/\lambda \right) / \lambda ~,
\label{coordrede}
\end{eqnarray}
where the conformal factor takes
the form (N.B.: $y^+_0 \equiv \sigma^+_0$)
\begin{eqnarray}
\rho = - {1 \over 2} \ln[1+{a \over \lambda}
e^{\lambda y^- + \lambda \Theta(y^+ - y^+_0) (y_0^+ - y^+)}]
{}~,
\label{rhoy}
\end{eqnarray}
and therefore the metric is
asymptotically constant on ${\cal I}_R^{\pm}$.

Using the fact that $T_{\mu \nu}$ transforms like a tensor under
diffeomorphisms, from (\ref{confstrozerosigma})
and (\ref{confstrommsigma})
one finds\cite{bhlect} that in the $y^\pm$ coordinates
\begin{eqnarray}
\left[T_{++}\right]_{{\cal I}^+_R} = 0 ~,~~~~
\left[T_{+-}\right]_{{\cal I}^+_R} = 0 ~,~~~~~~~~ ~~
\label{confstrozeros}\\
\left[T_{--}\right]_{{\cal I}^+_R} = { N \lambda^2 \over 48 \pi}
\left[ 1-{1\over\left(1+a e^{\lambda y^-}/\lambda\right)^2} \right]
{}~.
\label{confstromm}
\end{eqnarray}
Eq.(\ref{confstromm}) gives the flux of energy across ${\cal I}^+_R$.
Consistently with the picture of
black hole radiation\cite{haw75}, in the far past
of ${\cal I}^+_R$ ({\it i.e.} $y^- \rightarrow -\infty$) this
flux vanishes exponentially, while it
approaches a constant value as the
horizon ({\it i.e.} $y^- \rightarrow \infty$) is approached.

\subsection{Conformal-Gauge Analysis for Arbitrary $k$}
Let us now generalize the analysis to the case
in which the
matter energy-momentum tensor
satisfies the anomaly relations (\ref{anobw})
and (\ref{anobd}).
Since the anomalies are a quantum effect,
nothing changes concerning the black hole background metric,
but, instead of
(\ref{stroconfw}) and  (\ref{stroconfd}), the equations
satisfied by the
matter energy-momentum tensor in conformal gauge are now given by
\begin{eqnarray}
& T_{+-} = - {N \over 12 \pi} k^2 \partial_+ \partial_- \rho
{}~,&
\label{confw}\\
& \partial_\mp T_{\pm \pm} + \partial_\pm T_{+-}
- 2 T_{+ -} \partial_\pm \rho
= {N \over 12 \pi} k (1-k) \partial_\mp \partial_+ \partial_- \rho
{}~,&
\label{confd}
\end{eqnarray}
which lead to
\begin{eqnarray}
T_{\pm \pm} =
{N \over 12 \pi}
\left[ k \partial^2_\pm \rho - k^2 (\partial_\pm \rho)^2
+ t_\pm(\sigma^\pm) \right] ~.
\label{solconfd}
\end{eqnarray}
Obviously, (\ref{confw}), (\ref{confd}), and (\ref{solconfd})
reproduce (\ref{stroconfw}), (\ref{stroconfd}), and (\ref{solstroconfd})
when $k \! = \! 1$.

The functions of integration $t_\pm$  are to be determined
by requiring again that $T_{\mu \nu}$  vanish
on ${\cal I}_L^-$,
and that there be no incoming radiation
along ${\cal I}_R^-$ except
for the classical shock-wave at $\sigma^+ \! = \! \sigma^+_0$;
this leads again to $t_\pm \! = \! 0$.
Then using (\ref{confw}), (\ref{solconfd}),
and the expression of $\rho$ given
in (\ref{background}) we find that in
the $\sigma^\pm$ coordinates
\begin{eqnarray}
\left[\!T_{++}\! \right]_{{\cal I}^+_R}
= 0 ~,~~~~ \left[T_{+-} \right]_{{\cal I}^+_R} = 0 ~, ~~~~~~~
\label{confzerosigma}\\
\left[T_{--} \right]_{{\cal I}^+_R} =
{ N a \lambda^2 \over 48 \pi} k e^{\lambda \sigma^-}
{2 \lambda - k a e^{\lambda \sigma^-}
\over \left(\lambda - a e^{\lambda \sigma^-} \right)^2}
{}~.
\label{confmmsigma}
\end{eqnarray}

In order to get a clear
physical interpretation of this result we need to express it
in the $y^\pm$ coordinates like before.
In doing so, we shall take into account
the fact that, when $k \! \ne \! 1$,
$T_{\mu \nu}$ does not transform covariantly under coordinate
redefinitions of non-constant Jacobian.
For a conformal coordinate
redefinition $\sigma^+ \rightarrow y^+ \!
= \! \sigma^+$,$\sigma^- \rightarrow y^- \! = \! f(\sigma^-)$,
from
(\ref{deltamunu}) one finds that
$\Delta^{(\sigma,y)}_{++} \! = \! \Delta^{(\sigma,y)}_{+-} \! = \! 0$,
and
\begin{eqnarray}
\Delta^{(\sigma,y)}_{--} = {N \over 24 \pi}
(1-k) \left\{ \partial^2_- \ln({dy^- \over d\sigma^-})
- {(1-k) \over 2} [\partial_- \ln({dy^- \over d\sigma^-})]^2
- 2 k (\partial_- \rho) \partial_- \ln({dy^- \over d\sigma^-}) \right\}
{}~,
\label{delta}
\end{eqnarray}
which generalizes
the ordinary\cite{itz} ($k \! = \! 0$)
{\it Schwarzian derivative} of the conformal
map $\sigma \! \rightarrow \! y$,
to the case of our $k$-dependent
anomalous transformations of the
energy momentum tensor.

Since $d\sigma^-/dy^+ \! = \! 0$,
from (\ref{ttrasf}) and (\ref{confzerosigma}) it
follows that on ${\cal I}^+_R$
\begin{eqnarray}
\left[T^{(y)}_{++} \right]_{{\cal I}^+_R}
\! = \! 0 ~,~~
\left[T^{(y)}_{+-} \right]_{{\cal I}^+_R} \! = \! 0 ~,~~~~~~~~
\label{confzeros}\\
\left[T^{(y)}_{--}\right]_{{\cal I}^+_R}
\! = \! \left[(T^{(\sigma)}_{--} \! + \! \Delta^{(\sigma,y)}_{--})
(d \sigma^- / dy^-)^2 \right]_{{\cal I}^+_R}
{}~.
\label{confdeltaty}
\end{eqnarray}
and with a straightforward calculation we find that
\begin{eqnarray}
\left[T^{(\sigma)}_{--}
\left( {d \sigma^- \over
dy^-} \right)^2 \right]_{{\cal I}^+_R} \!\!\!\!&=&\!\!
{ N \lambda^2 \over 48 \pi} ~
{2k a e^{\lambda y^-}/\lambda +
(2k - k^2) a^2 e^{2 \lambda y^-}/\lambda^2
\over\left(1+a e^{\lambda y^-}/\lambda\right)^2}
{}~,
\label{tmmtrasfk}\\
\left[\Delta^{(\sigma,y)}_{--} \left( {d \sigma^- \over
dy^-} \right)^2 \right]_{{\cal I}^+_R} \!\!\!\!&=&\!\!
{ N \lambda^2 \over 48 \pi}
\left[ 1-{1 + 2k a e^{\lambda y^-}/\lambda +
(2k - k^2) a^2 e^{2 \lambda y^-}/\lambda^2
\over\left(1+a e^{\lambda y^-}/\lambda\right)^2} \right]
{}~.
\label{deltammk}
\end{eqnarray}
Adding these last two results
we see that the $k$-dependent terms cancel out,
and, obviously, the left-over formula
for $T_{--}$ on ${\cal I}^+_R$
exactly reproduces Eq.(\ref{confstromm}).
We conclude that the black hole
radiation observed in the $y^\pm$ coordinate system
is insensitive to the value of $k$.

Using the covariance of $T_{\mu \nu}$ discussed in Section 2,
we can deduce that the coordinate systems which can
be obtained from the $y^\pm$
coordinate system by a coordinate redefinition of Jacobian
asymptotically constant on ${\cal I}_R^{+}$
will also observe
$k$-independent black hole radiation.
Importantly, these are all the coordinate systems
with metric asymptotically constant on ${\cal I}_R^{+}$,
which obviously include all observers asymptotically
Minkowskian on ${\cal I}_R^{+}$.

\subsection{Light-Cone-Gauge Analysis for Arbitrary $k$}
We now want to show that also for the light-cone-gauge observers,
which we define as those
with $g_{--} \! = \! 0$ and $g_{+-} \! = \! - \! 1/2$,
the black hole radiation is $k$-independent.
Let us start by observing that the
anomaly relations (\ref{anobhatw}) and (\ref{anobhatd}) imply that
\begin{eqnarray}
\hat\nabla_\mu ({\hat g}^{\mu \nu} T_{\nu \alpha}) -
{1 \over 2} \partial_\alpha ({\hat g}^{\mu \nu} T_{\mu \nu})
= - {N \over 48 \pi} \partial_\alpha R({\hat g})
{}~,
\label{hardanomaly}
\end{eqnarray}
which does not depend explicitly on $k$;
it depends on $k$ only implicitly, through
the $k$-dependence of ${\hat g}_{\mu \nu}$.
This relation is particularly useful in light-cone gauge,
where
the metric $g_{\mu \nu}$
has constant determinant, and therefore the $k$-dependence
of ${\hat g}_{\mu \nu}$ is trivial.

Using (\ref{hardanomaly}) and (\ref{anobw}), one finds that
in light-cone gauge the matter energy-momentum tensor
satisfies the relations
\begin{eqnarray}
& T_{+-} + g_{++} T_{--} = - {N \over 24 \pi} k  \partial^2_- g_{++}
{}~,&
\label{lcw}\\
& \partial_- T_{++} + 2 \partial_- (g_{++} T_{+-})
- g_{++} \partial_+ T_{--}
= {N \over 24 \pi} \partial_+ \partial^2_- g_{++}
{}~,&
\label{lcdp}\\
& \partial_+ T_{--} + 2 \partial_- (g_{++} T_{--})
- g_{++} \partial_- T_{--}
= {N \over 24 \pi} \partial^3_- g_{++}
{}~.&
\label{lcdm}
\end{eqnarray}
Most importantly, the differential equation (\ref{lcdm})
involves only $T_{--}$ and is $k$-independent;
therefore, with $k$-independent boundary conditions,
it leads to $k$-independent $T_{--}$.
The general solution of (\ref{lcdm}) has the form
\begin{eqnarray}
T_{--} \! = \!\! - \! {N \over 24 \pi} \!
(\partial_- F(\xi^+ \! ,\!\xi^-))^2 \!
\left[ g_{++}(\xi^+ \! ,\!\xi^-) \{F(\xi^+ \! ,\!\xi^-),\xi^-\}
\! + \! {1 \over 2} \! \partial_- g^2_{++}(\xi^+ \! ,\!\xi^-)
\! + \! t^{lc}_-(F(\xi^+ \! ,\!\xi^-)) \right],
\label{solulcdm}
\end{eqnarray}
where $\xi^+$ and $\xi^-$ are light-cone coordinates,
$F$ is such that
\begin{eqnarray}
g_{++} = - {\partial_+ F \over \partial_- F}
{}~,
\label{gppf}
\end{eqnarray}
$\{,\}$ denotes the ordinary
{\it Schwarzian derivative}, and
$t^{lc}_-$, which is
a function of $\xi^+$ and $\xi^-$ only through $F$,
is to be fixed by imposing physical
boundary conditions.

It is easy to verify explicitly that, for light-cone-gauge observers,
the black hole radiation is insensitive to the value of $k$.
In light-cone gauge
the black hole background metric that
we have been considering can be described by (N.B.: the shock-wave
is at $\xi^+ \! = \! \xi^+_0$)
\begin{eqnarray}
g_{++} = 1 + \Theta(\xi^+ - \xi^+_0)
\left[ a e^{\lambda (\xi^- - \xi^+ + \xi^+_0)}/\lambda - 1 \right]
{}~,
\label{gppbh}
\end{eqnarray}
which corresponds to
\begin{eqnarray}
F = - {1 \over \lambda} \ln \left[ e^{\lambda (\xi^+ - \xi^- - \xi^+_0)}
- a/\lambda \right] + \Theta(\xi^+ - \xi^+_0) \, (\xi^+ - \xi^+_0)
{}~.
\label{fbh}
\end{eqnarray}
We observe that this $F$ also has a geometrical interpretation; in fact,
the $\xi^\pm$ coordinates that we are using in light-cone gauge
and the $y^\pm$ coordinates that we used in the preceding
subsections are related by
\begin{eqnarray}
y^+ = \xi^+ ~,~~~~y^- = F(\xi^+,\xi^-)
{}~.
\label{xymap}
\end{eqnarray}

The physical boundary conditions needed to fix $t^{lc}_-(\xi^+,\xi^-)$
are again\footnote{Since the $\sigma^\pm$ coordinates
are asymptotically Minkowskian on ${\cal I}_L^-$ and ${\cal I}_R^-$
(see Eq.(\ref{background})),
there is a coordinate redefinition of Jacobian
asymptotically constant on ${\cal I}_L^-$ and ${\cal I}_R^-$
which connects the $\sigma^\pm$ coordinate system
and any given light-cone gauge coordinate system.
Therefore, in order to obtain the corresponding
boundary conditions in a given light-cone gauge coordinate system,
we can transform covariantly the boundary conditions
for $T_{\mu \nu}$ imposed on ${\cal I}_L^-$ and ${\cal I}_R^-$
in the $\sigma^\pm$ coordinate system.}
provided by the requirement
that $T_{\mu \nu}$  vanish
on ${\cal I}_L^-$
and that there be no incoming radiation
along ${\cal I}_R^-$ except
for the classical shock-wave at $\xi^+ \! = \! \xi^+_0$.
This leads to
\begin{eqnarray}
t^{lc}_- = {\lambda^2 \over 2}
\left[ 1-{1\over\left(1
+ a e^{\lambda F(\xi^+,\xi^-)}/\lambda\right)^2} \right]
{}~.
\label{tmlc}
\end{eqnarray}
Eqs.(\ref{solulcdm}), (\ref{gppbh}), (\ref{fbh}), and (\ref{tmlc})
completely determine $T_{--}$, and in particular
on ${\cal I}_R^{+}$ one finds that
\begin{eqnarray}
\left[T_{--}\right]_{{\cal I}^+_R} = { N \lambda^2 \over 48 \pi}
\left[ 1-{1\over\left(1+a e^{\lambda \xi^-}/\lambda\right)^2} \right]
{}~,
\label{lcmm}
\end{eqnarray}
which, as expected, indicates that the black hole radiation
observed in the $\xi^\pm$ coordinates
is insensitive to the value of $k$.
Since any
two light-cone-gauge observers are connected by a diffeomorphism
of unit Jacobian, which is a symmetry of the theory for any $k$,
the $k$-independence of the black hole radiation
observed in the $\xi^\pm$ coordinates
also applies to any other light-cone-gauge observer.

Note that (\ref{lcmm}) is identical to (\ref{confstromm}).
This is due to the fact that, as shown by (\ref{gppbh}),
also in the $\xi^\pm$ coordinates the metric is
asymptotically constant on ${\cal I}_R^{+}$, and,
as shown by (\ref{xymap}), the map between $\xi^+$,$\xi^-$
and $y^+$,$y^-$ is the identity on ${\cal I}_R^{+}$.

For completeness we also notice
that, having solved for $T_{--}$ and fixed
the above mentioned boundary conditions,
one can use (\ref{lcw}) and (\ref{lcdp})
to derive $T_{++}$ and $T_{+-}$, and in particular
on ${\cal I}_R^{+}$ one finds again that
\begin{eqnarray}
\left[T_{++}\right]_{{\cal I}^+_R} = 0 ~,~~~~
\left[T_{+-}\right]_{{\cal I}^+_R} = 0 ~.
\label{lczeros}
\end{eqnarray}

\section{Conclusion}
To summarize, in our
semiclassical analysis of black hole radiation in matter-coupled
dilaton gravity, we have considered
a one-parameter $k$-family of measures for the path integral
quantization of the matter fields.
We have derived several symmetry properties of these measures,
including a formula for the non-covariant
transformation
of the matter energy-momentum tensor
under coordinate redefinitions,
and observed that
the Weyl anomaly is proportional to the parameter $k$.
We have found that all these
quantizations of the matter fields
are consistent with the phenomena of black hole radiation,
and that the radiation seen by
all observers whose metric is
asymptotically constant on ${\cal I}_R^{+}$, which are the
observers ordinarily used in the description of
black hole radiation,
is insensitive to the value of $k$.
We have verified explicitly
this $k$-independence for two such observers,
one in conformal gauge
and the other in light-cone gauge,
and used the covariant conservation of the matter
energy-momentum tensor under coordinate redefinitions
of constant Jacobian to deduce its
validity for any other such observer.

Our results should also clarify the relation between anomalies
and black hole radiation in 1+1 dimensions.
The usual claim that the black hole radiation is a consequence
of the Weyl anomaly, should be understood as strongly
dependent on the assumption that the matter energy-momentum
tensor be covariantly conserved at the quantum level.
In general, the presence of any (Weyl and/or diffeomorphism)
anomaly is sufficient to support black hole radiation.
The insensitivity of black-hole radiation
to the parameter characterizing
the measure
must be understood as a feature of the particular (bosonic)
theory that we considered; in fact, based on the experience with
the chiral Schwinger model\cite{chirschw}
(where a
one-parameter ``$a$"-family
of chiral symmetry breaking measures
has been investigated, and the
mass emergent in that theory
does depend on $a$),
one can expect that
parameters characterizing the measure have a non-trivial
physical role
in more general theories (particularly when gravity is coupled to
chiral matter\cite{ags,chiranom}).

Interestingly, in our light-cone-gauge analysis
a key role was played by
the relation (\ref{hardanomaly}),
which in every gauge depends only implicitly on $k$ and
in light-cone gauge is completely $k$-independent.
This relation generalizes the one ($k \! = \! 0$) encountered
in Ref.\cite{abs95} to the case of the $k$-dependent
anomalies (\ref{anobw}),(\ref{anobd}).
The results found in the present paper agree with the
expectation\cite{abs95} that this relation encodes some essential
feature of the theory.

We also observe that the singularity of the
limit $k \! \rightarrow \! 0$ in Eq.(\ref{ktrasf})
was not encountered in any of the results which have followed.
Further investigation of the possible consequences
of this singularity would be interesting.
It is plausible that it may surface as a non-analyticity to be handled
in the higher orders of the semiclassical approximation,
but it may also turn out to be simply an accidental result
of the type of parametrization that we have chosen.

\bigskip
\bigskip
We thank E. Keski-Vakkuri for suggesting that
the results of Ref.\cite{abs95} might be important for the
understanding of the relation between Weyl anomaly and black hole
radiation, and
L. Griguolo and R. Jackiw for
very useful comments.
\bigskip
\bigskip

\vglue 0.6cm
\centerline{\Large {\bf Note Added}}
\vglue 0.3cm

Upon completion of our manuscript, L. Griguolo
brought to our attention Ref.\cite{navarri}, in which
1+1-dimensional
black hole radiation is analyzed semiclassically
in conformal gauge
assuming that the matter energy-momentum tensor
be traceless, but not covariantly conserved,
{\it i.e.} the special case $k \! = \! 0$
in the one-parameter $k$-family of quantizations that we considered
here.

\newpage
\baselineskip 12pt plus .5pt minus .5pt

\end{document}